\documentstyle[aps,preprint]{revtex}
\newlength{\defaultparindent}
\newenvironment{Default Paragraph Font}{}{}

\begin{document}
\draft
\title{Grain Boundary Induced Magneto-Far Infrared Resonances in Superconducting YBa%
$_2$Cu$_3$O$_{7-\delta }$ Thin Films}
\author{H.-T. S. Lihn, E.-J. Choi, S. Kaplan, H. D. Drew}
\address{Center for Superconductivity Research, Physics Department, University of\\
Maryland, College Park, MD 20742\\
and\\
Laboratory for Physical Sciences, College Park, MD 20740}
\author{Q. Li and D. B. Fenner}
\address{Advanced Fuel Research, East Hartford, Connecticut 06138}
\date{\today }
\maketitle

\begin{abstract}
Spectral features induced by 45$^{\circ }$ in-plane misoriented grains have
been observed in the far infrared magneto-transmission of YBa$_2$Cu$_3$O$%
_{7-\delta }$ thin films. Two strong dispersive features are found at 80 and
160 $cm^{-1}$ and a weaker one at 116 $cm^{-1}$. The data can be well
represented by Lorentzian oscillator contributions to the conductivity.
Several possible interpretations are discussed. We conclude that the
resonances are due to vortex core excitations.\\ \\ 
\end{abstract}


\narrowtext
The electronic properties of superconductors are strongly affected by the
application of magnetic fields. The effects range from the quantum
interference in Josephson junctions to dissipation in current carrying
wires. These effects present interesting physical questions as well as
important challenges for superconductivity applications. For example the
Hall coefficient is generally observed to reverse sign in the
superconducting state, and in some cases even reverse back before vanishing
at temperatures well below T$_c$ due to pinning of the vortex lattice\cite
{Lobb}. This behavior is known to be a consequence of the complex dynamics
of vortex motion in the presence of viscous and pinning forces but is not
yet well understood\cite{Dorsey,Ferrell}. Indeed controversies about such
fundamental issues as the vortex effective mass\cite{Suhl} and the magnetic
force on the moving vortex\cite{Dorsey} remain unresolved after decades of
study of vortex dynamics. The early theories were essentially
phenomenological but recently efforts have been devoted to the development
of theories of vortex dynamics based on microscopic physics\cite{Hsu,Kopnin}%
. The phenomenological models have been extensively employed in describing
the microwave losses\cite{Clem}. Since many of the characteristic
frequencies of the vortex system lie above microwave frequencies, recent
high frequency measurements on high T$_c$ superconductors have greatly
expanded the available phenomenology of this subject. Pulsed terahertz
experiments on YBa$_2$Cu$_3$O$_{7-\delta }$ (YBCO) have shown a strongly
temperature dependent ac Hall signal that has been interpreted in terms of
the response of thermally excited quasiparticles near the nodes of a d-wave
gap in YBCO\cite{Orenstein}. At higher frequencies (but below the IR gap)
two chiral resonances have been reported above which the system exhibits
free hole-like optical activity\cite{Karrai,Choi}. The chiral resonances
have been identified as the pinning resonance and the vortex core resonance
within a clean limit theory of vortex dynamics developed by T. C. Hsu\cite
{Hsu}. In this picture the bare resonances are hybridized with the
underlying cyclotron resonance of the holes and they are observable only
because of the presence of the vortex pinning.

In this paper we present an investigation of the far infrared (FIR)
magneto-optics of YBCO films containing large-angle misoriented in-plane
grains. The grain boundaries provide a reduced symmetry for pinned vortices
which can be expected to alter the selection rules for optical transitions
in the vortex core. However, grain boundaries are also known to modify the
electronic properties of superconductors even in zero magnetic field. The
microstructure of various types of grain boundaries in YBCO films have been
studied by electron energy loss spectroscopy (EELS) together with
transmission electron microscopy (TEM)\cite{Zhu,Browning,Chisholm}. Hole
depletion is found that extends as far as 60\AA\ from asymmetric grain
boundaries, whereas for symmetric grain boundaries the hole density remains
essentially constant. They can act as weak link Josephson junctions and
enhance the microwave losses\cite{Miller,Laderman}. In long junctions
standing wave Josephson modes can be excited leading to Fiske steps in the
I-V characteristics\cite{Winkler}. Also, spectral features in the tunneling
conductance have been reported on a single grain boundary\cite{Chaudhari}.
Both of these effects are strongly perturbed by the application of a
magnetic fields\cite{Winkler,Chaudhari}.

We report far infrared magneto-optical transmission measurements on YBCO
films with varying densities of 45$^{\circ }$ oriented grains\cite{gb45}.
Induced resonant spectral features are observed superposed on
magneto-optical features observed on samples with low densities of 45$%
^{\circ }$ grains reported earlier and discussed in terms of vortex dynamics.%
\cite{Choi,Vortex} The magnetic field dependence and temperature dependence
of the spectra are studied and the spectral line shapes of the induced
features are analyzed using a Lorentzian oscillator curve fitting. The
possible origin of these resonances is discussed.

The samples are YBCO thin films grown by Pulsed Laser Deposition (PLD) on
silicon substrates with yittria stabilized zirconia (YSZ) buffer layers and
cap layers. The film thickness is typically 400\AA\ and typical critical
temperatures are T$_c$ = 89$\pm 1^{\circ }$K measured by ac susceptibility.
Patterned line critical current densities are J$_c\approx 2*10^6$A/cm$^2$ at
77$^{\circ }$K\cite{Fork}. X-ray rocking curve measurements show that c-axis
alignment is typically within 0.7$^{\circ }$, which is comparable with
samples grown on LaAlO$_3$ substrates. The morphology of the films is
studied by X-ray diffraction. X-ray $\phi $ scans performed on the (205)
plane show that the samples fall into two classes: (1) Films having
appreciable densities of low angle grains but no measurable high angle
grains. (2) Films showing a high concentration of high angle (mainly 45$%
^{\circ }$) in-plane grains, but little or no low angle grains. In the
samples from Class (2), the 45$^{\circ }$ signal is as large as 4\% relative
to the signal from the grains aligned with the substrate. Although these
samples were not especially grown for misoriented grains it is known that
the density of 45$^{\circ }$ grains increases with the growth temperature.

The transmission of the films is measured with a fast scan FTIR\
spectrometer using a 2.2$^{\circ }$K bolometer detector. External magnetic
fields up to 12T are applied perpendicular to the a-b plane of the YBCO thin
films and the sample temperature is varied from 2.2$^{\circ }$K to above T$_c
$. The incident FIR radiation is elliptically polarized by a polarizer
comprised of a metal grid linear polarizer and a 0.9mm thick x-cut quartz
waveplate placed in front of the samples. The peak efficiency of the
polarizer for circular polarization is at 65 $cm^{-1}$ and 160 $cm^{-1}$.
The transmission coefficient is related to the conductivity by $T^{\pm
}(H)=4n\left/ \left| \left( Z_0d\sigma ^{\pm }(H)+n+1\right) \right|
^2\right. $ where $\sigma ^{+(-)}(H)$ is the circularly polarized
conductivity in hole cyclotron resonance (hCR) active (inactive, or eCR)
mode, $d$ is the thickness of the YBCO film, $Z_0$ is the free space
impedance and $n$ is the refractive index of the silicon substrate. The
elliptically polarized transmission data was transformed to the circular
polarized response, $T^{\pm }(H)$, using a calibration of the polarizer
efficiency based on the cyclotron resonance of a two dimensional electron
gas in a high mobility GaAs quantum well\cite{Karrai,Wu}. Since the
efficiency of the polarizer is a slowly varying function over a wide
frequency range, the unfolding introduces little error to the spectral
features except when the polarizer is close to the half waveplate condition
at about 130 $cm^{-1}$. A more detailed description of the experiment and
data manipulation is given elsewhere\cite{Karrai,Choi,Vortex}.

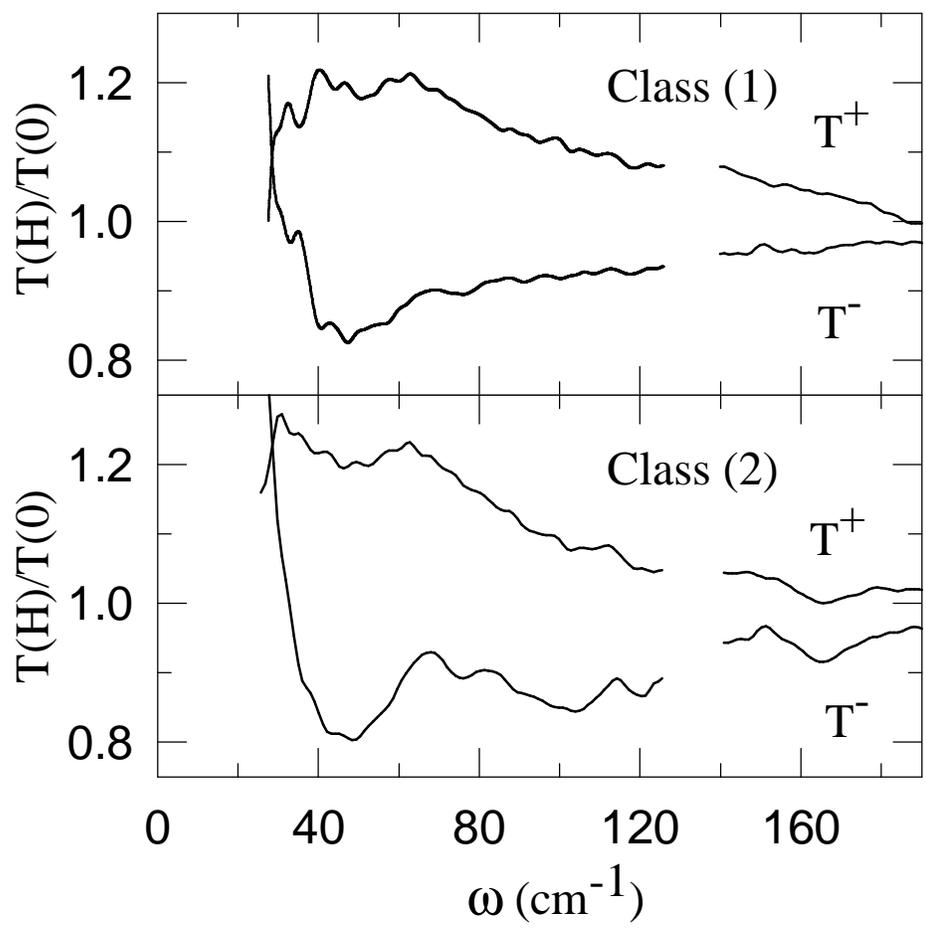
\begin{figure}[tbp]
\caption{ The transmission ratio $T^{\pm }(H)/T(0)$ vs. frequency for YBCO
thin films from Class (1) and (2) at 12T and 2.2$^{\circ }$K. Class(1): YBCO
on Si with only low angle grains. Class (2): YBCO on Si substrate with 45$%
^{\circ }$ grains. The circularly polarized response is shown: $T^{+}$ for
the hCR mode and $T^{-}$ for the eCR mode. The region between 125$\sim $140 $%
cm^{-1}$ corresponds to a quartz phonon and the half waveplate condition of
the polarizer. The signal to noise ratio deteriorates rapidly below 40 $%
cm^{-1}$ because of low transmitted spectral intensity. }
\label{t12}
\end{figure}

The transmission ratio $T^{\pm }(H)/T(0)$ of samples from Class (1) and (2)
at 12T and 2.2$^{\circ }$K are shown in Figure \ref{t12}. The lower (upper)
curves $T^{+(-)}(H)/T(0)$ are in hCR (eCR) mode. The measurement errors
become large below 40 $cm^{-1}$ due to the low transmitted intensity and in
the region between 125$\sim $140 $cm^{-1}$ due to a quartz phonon and the
half waveplate condition of the polarizer. The transmission curves of Class
(1) samples are generally smooth except a for a small broad peak feature at
65 $cm^{-1}$ in the eCR mode and a sharp rise below 50 $cm^{-1}$ in the hCR
mode. In addition, the transmission ratio approaches unity approximately as $%
\omega ^{-1}$ at high frequencies in both modes. These features have been
observed in many high quality YBCO samples and they are qualitatively
independent of the detailed sample quality, growth technique and choice of
substrates. As can be seen from the figure these spectral features are also
present in Class (2) samples. We have attributed these canonical features to
the electrodynamic response of pinned vortices\cite{Choi,Vortex}.

The conventional theories of vortex dynamics do not properly describe the
far infrared magneto-optical response of class (1) samples\cite{Clem}. In
these theories the conductivity has only one resonance and it is at zero
frequency with a width that depends on vortex viscosity $\eta $. In
particular, the optical activity $T^{+}(H)/T^{-}(H)$ observed in the FIR\
experiments is absent. A recently developed clean limit theory of vortex
dynamics by T. C. Hsu does successfully describe the FIR experiments on
class (1) samples \cite{Hsu,Choi}. In Hsu's picture, the small feature in $%
T^{+}(H)/T(0)$ at 65 $cm^{-1}$ is due to the hybridized vortex core
resonance and the sharp rise below 50 $cm^{-1}$ in $T^{-}(H)/T(0)$ is the
high frequency part of the hybridized pinning resonance. The experiments do
not measure down to sufficiently low frequencies to fully resolve this
pinning resonance\cite{lowfreq}. The high frequency optical activity is a
consequence of these two chiral resonances.

Comparing the spectra of the two classes of samples, several differences are
observed and these differences are the subject of this paper. There are
several induced dispersive features, centered around 80, 116, and 160 $%
cm^{-1}$, superimposed on spectra of the class (1) form. This distribution
suggests a periodicity of $\sim $ 40 $cm^{-1}$. In addition, the feature at
80 $cm^{-1}$, which occurs in the hCR mode only, is rather wide extending
about 40 $cm^{-1}$, and in fact it appears that it may be a doublet
consisting of features at 72 $cm^{-1}$ and 90 $cm^{-1}$. The features at 116 
$cm^{-1}$ and 160 $cm^{-1}$ occur with nearly equal weight in both modes.
The 116 $cm^{-1}$ feature is quite narrow, about 10 $cm^{-1}$ wide compared
with a width of about 20 $cm^{-1}$ for the 160 $cm^{-1}$ feature.

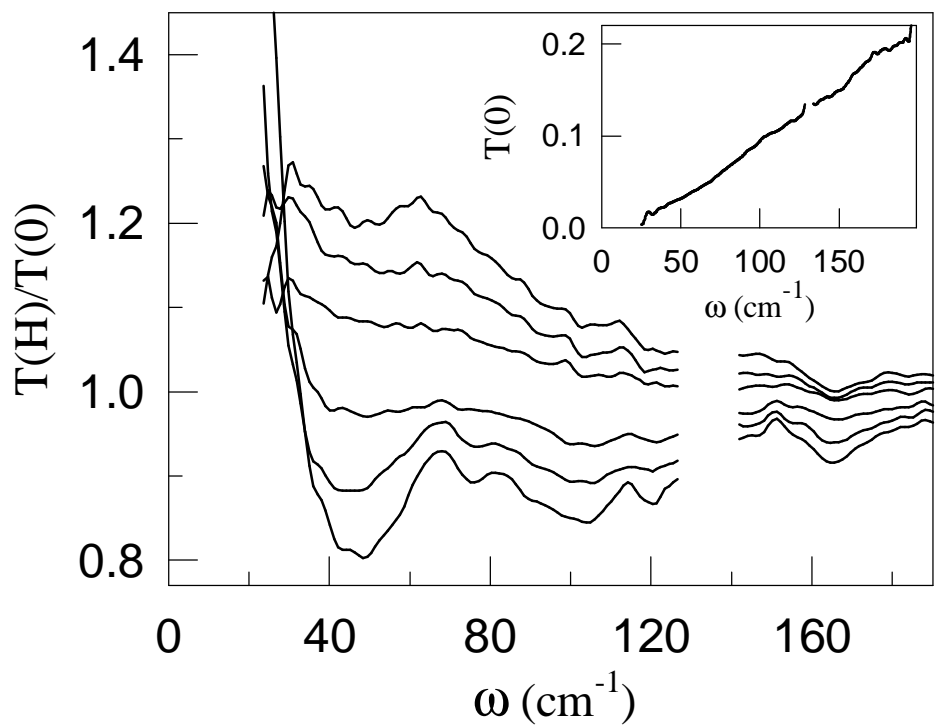
\begin{figure}[tbp]
\caption{ The magnetic field dependence of $T^{\pm }(H)/T(0)$ of a YBCO thin
film of Class (2) at 2.2$^{\circ }$K. Fields from top are +12T, +8T, +4T,
-4T, -8T, -12T. Several features induced by the 45$^{\circ }$ grain
boundaries become distinct at high fields. There is a dispersive spectral
feature centered at 80 $cm^{-1}$ in the hCR mode and features at 116 $cm^{-1}
$ and 160 $cm^{-1} $ occur with equal weight in both modes. These features
do not shift as field increases and among different Class (2) samples.
Inset: Transmission of a class (2) sample at zero field and 4$^{\circ }$K.
This smooth response corresponds to a simple London-Drude conductivity. }
\label{fielddep}
\end{figure}

The magnetic field dependence of $T^{\pm }(H)/T(0)$ of a Class (2) sample is
shown in Figure \ref{fielddep}. The amplitudes of the grain boundary induced
features in Figure \ref{t12} are seen to grow approximately linearly with
magnetic field and there is no measurable shift in their frequency
positions. This observation suggests an interpretation in terms of vortex
dynamics since the vortex density is proportional to the magnetic field and
since the energy spacings of the vortex core levels are not expected to
change significantly for magnetic fields small compared to $H_{c2}$.
Although these features are induced by the 45$^{\circ }$ grain boundaries,
they do not appear to be strongly affected by any disorder associated with
the grains. A comparison of the spectra from many similar samples shows that
the positions of these induced features are very reproducible from sample to
sample and their amplitudes correlate with the strength of the 45$^{\circ }$
peak in the X-ray $\phi $ scans.

We suggest that these induced spectral features are due to vortex core
excitations. According to T. C. Hsu the dipole transitions between the
quasiparticle levels in the vortex core are suppressed when the system is
translationally invariant\cite{Hsu,Drew:Hsu}. The addition of a symmetric
pinning potential breaks the translational symmetry and induces the angular
momentum conserving lowest vortex core transition. However, for vortices
pinned at a grain boundary the rotational symmetry is also lost allowing, in
principle, violation of the angular momentum selection rule and the higher
energy transitions\cite{Zhu:Drew,Janko}.

For the interpretation in terms of vortex dynamics we consider two
scenarios. In the first we take the feature at 65 $cm^{-1}$ in $T^{+}(H)/T(0)
$ as the fundamental of the vortex core resonance. This follows from
extensive studies of the magneto-optical properties of Class (1) films which
also show this feature\cite{Choi,Vortex}. Within the Hsu model the vortex
core resonance is hybridized with the pinning resonance. The corresponding
core level spacing $%
\hbar
\Omega _0=E_{1/2}-E_{-1/2}$ is about 40 $cm^{-1}$ (where $\mu =\pm 1/2$ are
the angular momentum quantum number of vortex core levels). The higher
frequency resonances are then the $\Delta \mu >1$ transitions induced by the
grain boundaries. In s-wave BCS theory these resonances occur at near
multiples of $\Omega _0$ if $%
\hbar
\omega \ll 2\Delta $. In this scenario the resonance at 80 $cm^{-1}$ is the
second harmonic, corresponding to the transition from $\mu =-1/2$ to $\mu
=+3/2$ or $\mu =-3/2$ to $\mu =+1/2$. As one can see from the curve fitting
described later, this 80 $cm^{-1}$ feature appears to be split into a
doublet. This splitting may arise from the lifting of the degeneracy of the
two transitions due to the grain boundary. The dispersive features at 116 $%
cm^{-1}$ and 160 $cm^{-1}$, occur in both modes with nearly equal strength.
We conjecture that these features are the third and fourth harmonics of the
vortex core resonance. However, a splitting such as observed on the second
harmonic may also play a role in these transitions, for example, they could
be the split third harmonic.

In the second scenario, we regard the 80 $cm^{-1}$ feature as the
fundamental resonance of the vortex core, the transition from $\mu =-1/2$ to 
$\mu =+1/2$. According to the dipole selection rules, this resonance should
occur in the hCR mode, as observed\cite{Zhu:Drew}. In this case the
splitting must come from two different types of sites for the vortices on
the grain boundaries. Then the features at 116 $cm^{-1}$ and 160 $cm^{-1}$
are the split second harmonic. However, this interpretation seems to
produces a paradox. If the 65 $cm^{-1}$ feature in the eCR mode in both
Classes of samples is also related to the fundamental of the vortex core
resonance, why does the resonance occur at two different frequencies? This
problem may be resolved by depolarization shifting of the allowed resonance.
Zhu {\it et al}.\cite{Zhu:Drew} have examined the depolarization effect on
the response due to the inhomogeneous conductivity of the vortex system. The
resonance frequency $2E_{1/2}$ can be redshifted to $2(1-\kappa )E_{1/2}$ in
which $\kappa $ is estimated to be 0.43. This would imply, for the dipole
allowed transition, that those vortices in the bulk participate in the
response as described by Hsu but are subject to the depolarization effect,
while those vortices pinned by 45$^{\circ }$ grain boundary give rise to a
weakly allowed resonance that is not depolarization shifted. A possible
difficulty with this interpretation is that the dipole resonance is quenched
by the vortex motion, according to Hsu, so that the depolarization shift
may, in fact, be small.

The temperature dependence of the magneto-optical spectra from a Class (2)
samples is shown in Figure \ref{tempdep}. The unpolarized transmission ratio 
$T(9T)/T(0)$ is shown for a sample similar to the one used for the low
temperature data in Figure 1. In these spectra the resonance at 80 $cm^{-1}$
is not as well resolved because the grain boundary induced features in this
sample are weaker and they are mixed with the low frequency rise of the
pinning resonance. However, we see that the structures at 110 $cm^{-1}$ and
160 $cm^{-1}$ are present and they are seen to persist up to very high
temperatures, remaining discernible at 55$^{\circ }$K. Moreover, the
frequency positions of these features do not shift with temperature. This
behavior is very different than predictions for vortex core excitations
within the BCS s-wave theory. In this case calculations have shown that when
the temperature is comparable with $E_{1/2}/k_B$ which is about 20$^{\circ }$%
K (40$^{\circ }$K) for $\Omega _0$= 40 $cm^{-1}$ (80 $cm^{-1}$), the $\mu
=1/2$ level becomes thermally populated with quasiparticles and the
self-consistent gap function begins to change its shape. The vortex core
broadens and the quasiparticle energy level spacings decrease. The observed
behavior would require the gap function in YBCO to be nearly temperature
independent as has also been reported from infrared reflectivity measurements%
\cite{Collins}. Also, we note that for the case of d-wave superconductors
the quasiparticle levels in the vortex core have a number of significant
differences from the s-wave case. In addition to the set of localized levels
similar to that found in s-wave superconductors there are also continuum
levels and additional high energy levels outside the core that are
associated with s-wave admixture induced by the vortex\cite{Soininen,Ren}.
Therefore, another possibility is that the high frequency features are
related to these levels.

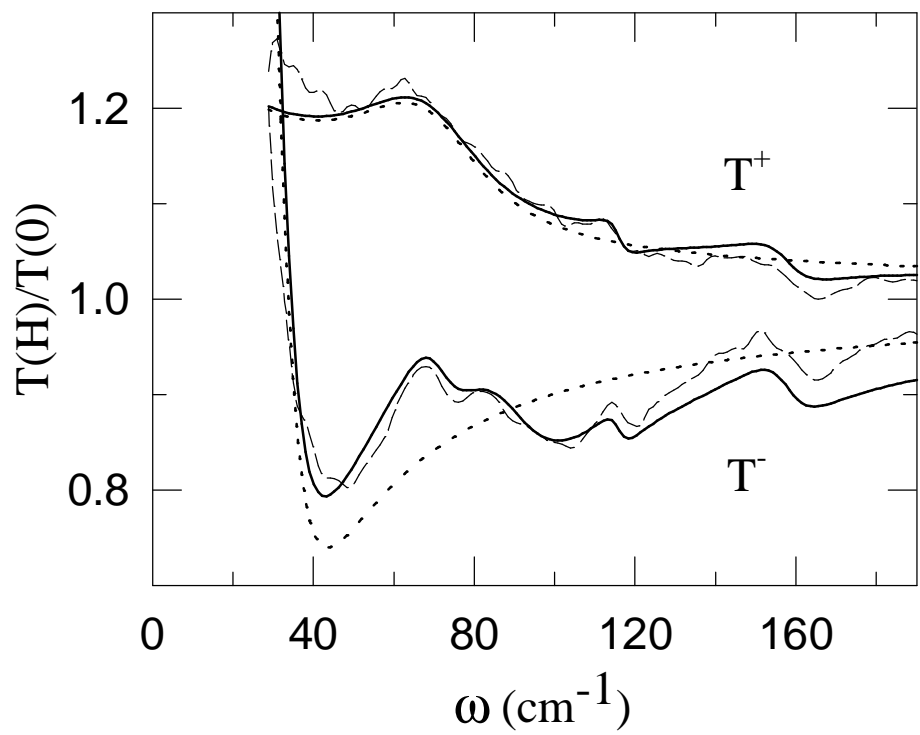
\begin{figure}[tbp]
\caption{ The results of the line shape analysis of Class (2)
magneto-transmission data using Hsu's conductivity plus additional
Lorentzian oscillators. The long dashed lines are $T^{\pm }(H)/T(0)$ of YBCO
thin film from a Class (2) sample in 12T field at 2.2$^{\circ }$K. The
dotted lines are the fits using only Hsu's vortex dynamics theory. The thick
solid lines are the fits including multiple Lorentzian oscillators to model
the vortex core excitations. The parameters for the fits are listed in Table 
\ref{tabfit}. }
\label{fit}
\end{figure}

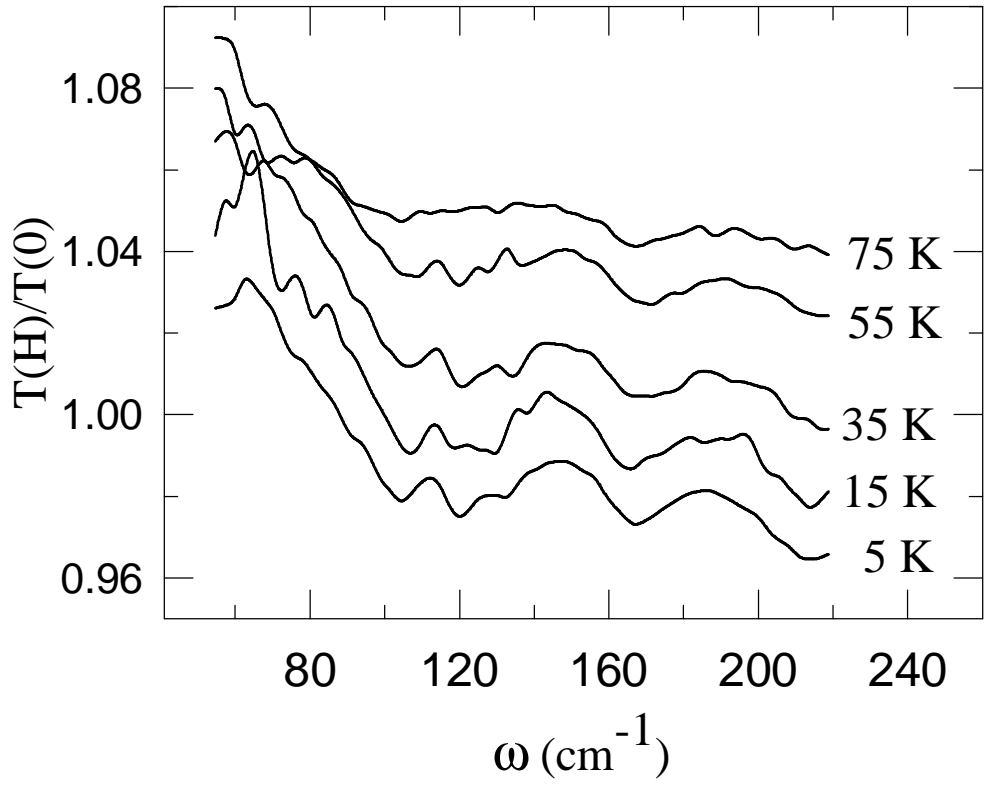
\begin{figure}[tbp]
\caption{ The temperature dependence of the unpolarized transmission ratio $%
T(H)/T(0)$ of a YBCO thin film of Class (2) at 9T. The dispersive features
at 116 $cm^{-1}$ and 160 $cm^{-1}$ persist to 55$^{\circ }$K. Because the
grain boundary related features in this sample are weaker the changes of the
features at 80 $cm^{-1}$ are not well resolved. }
\label{tempdep}
\end{figure}

To gain further insight into the nature of these grain boundary induced
optical features we have studied their spectral line shapes. We have modeled
the conductivity function of the system in terms of Lorentzian oscillators.
We start with the analysis of the spectra of the class (1) samples. These
spectra can be modeled in terms of Hsu's conductivity function\cite{Hsu,Choi}
or equivalently as two finite frequency oscillators and a zero frequency
delta function. To model the class (2) samples we first fit the 12T data in
Figure \ref{fielddep} to the Hsu conductivity or its Lorentzian equivalent.
The resulting Hsu fitting parameters are $\omega _c=3.7cm^{-1}$, $\Omega
_0=45cm^{-1}$, $\alpha =48cm^{-1}$, $1/\tau _\alpha =31cm^{-1}$. We have
then added Lorentzian oscillators to this conductivity function to model the
grain boundary induced dispersive features. We have included four new
oscillators to model the system 
\begin{equation}
\sigma _{total}^{\pm }=(1-\sum_{i=1}^4f_i^{\pm })\sigma _{Hsu}^{\pm
}+\sum_{i=1}^4f_i^{\pm }\sigma _i  \label{total}
\end{equation}
in which 
\begin{equation}
\sigma _i=\frac{ne^2}m\frac 1{i(\omega -\omega _i)+1/\tau _i}  \label{sigma}
\end{equation}
where $f_i^{\pm }$ represents the strength of the $i$th oscillator in two
polarization modes. The parameters from the fit are shown in Table \ref
{tabfit}. The results of these fits are shown in Figure \ref{fit}. We have
assumed that there is no contribution in the eCR mode from the first two
oscillators near 80 $cm^{-1}$(That is, $f_1^{+}=f_2^{+}=0$) because we
believe that this resonance is already included in $\sigma _{Hsu}^{+}$. The
third and fourth oscillators do not have an obvious chirality and their
oscillator strengths are an order of magnitude smaller than the first two.
This analysis shows that the dispersive line shapes of the grain boundary
induced spectral features in the transmission are consistent with resonances
at the center frequencies of these features.

It is interesting to consider the relation of these spectral features to the
electronic properties of grain boundaries. Numerous studies have been made
of the I-V characteristics of junctions associated with grain boundaries in
YBCO films grown on bicrystals\cite{Winkler,Chaudhari}. Winkler {\it et al.}%
\cite{Winkler} reported the observation of Fiske steps at low voltages due
to terahertz electromagnetic resonances in grain boundary junctions. These
are Josephson effect standing waves in the junction and they are suppressed
by a field of a few tenths of gauss. The characteristic field for the
suppression, $\phi _0\left/ (\lambda _LL)\right. $ (where $\phi _0$ is the
flux quantum, $\lambda _L$ London penetration depth, $L$ length of the grain
boundary), corresponds to the condition that the junction contains one flux
quantum. Chaudhari {\it et al.}\cite{Chaudhari} reported periodic structure
in the conductance of similar grain boundaries at higher voltages whose
field dependence is on a much larger scale. Conductance peaks are found at
about 2.4, 4.7, 7.2, and 9.5 mV and they have a very peculiar magnetic field
dependence. They find that the conductance peak at 9.5meV is suppressed when
a parallel magnetic field is applied (at about 2T) and is shifted when a
perpendicular field is applied. These effects are not understood but the
authors suggest that they may be related to quasiparticle levels in the
junction. From BCS calculations De Gennes {\it et al.}\cite{DeGennes}
predicted quasiparticle levels below the bulk gap in S-N junction that
depend on the details of the junction.

However, it does not appear to be possible to relate these effects to our
FIR results. This would require that the observed spectral features be
present in zero field and become suppressed in applied magnetic fields.
However, we do not observe any such features of the required strength in our
zero field transmission spectra as can be seen in the inset of Figure \ref
{fielddep}. We assume, for example, that there is a zero field structure of
30\% which diminishes linearly with field, it would give rise to a 10\%
change in transmission at 10T (See Fig. \ref{fielddep}) and go to zero at $%
\simeq $ 30T. For the Josephson standing waves this would imply a junction
of length less than 10\AA\ which is unreasonably small. In general it is
difficult to obtain a characteristic field of 30T since $H_{c2}$ $\sim $%
150T. Also, we note that the reproducibility of the observed spectral
features would require a reproducible ordered morphology of the grain
boundaries.

From these considerations we believe that these grain boundary related far
infrared features are induced by the magnetic field. Therefore it appears
that the picture in which these features are vortex core excitations brought
about by vortex pinning at 45$^{\circ }$ grain boundaries is the most
plausible interpretation of these magneto-optical experiments. Because the
80 $cm^{-1}$ feature in the hCR mode has the correct selection rule for the
dipole transition, we believe it represents the fundamental $\Delta \mu =1$
vortex core resonance. In this case the core spacing, as perturbed by the
grain boundary, $%
\hbar
\Omega _0\simeq $ 10meV. A calculation of the depolarization shifted
response of vortices in the bulk would help clarify the interpretation.

We acknowledge useful discussions with T. Hsu, C. Kallin, and F.-C. Zhang.
This work is partially supported by NFS under Grant No. 9223217.

\begin{table}[tbp]
\centering
\begin{tabular}{ccccc}
fitting$\left\backslash i\text{th}\right. $ & 1st & 2nd & 3rd & 4th \\ 
\tableline $\omega _i$ & 72$cm^{-1}$ & 90$cm^{-1}$ & 116$cm^{-1}$ & 158$%
cm^{-1}$ \\ 
$\frac 1{\tau _i}$ & 8$cm^{-1}$ & 16$cm^{-1}$ & 3.5$cm^{-1}$ & 8$cm^{-1}$ \\ 
$f_i^{+}$ & 0\% & 0\% & 0.04\% & 0.06\% \\ 
$f_i^{-}$ & 0.5\% & 1.3\% & 0.06\% & 0.15\%
\end{tabular}
\caption{ The parameters of the line shape analysis of the transmission data
using the Hsu conductivity and multiple Lorentzian oscillators. The formula
for the oscillators is given by Eq.(\ref{total}) and Eq.(\ref{sigma}). See
also Fig.\ref{fit}. }
\label{tabfit}
\end{table}


\begin{references}
\bibitem{lowfreq}  Recently the pinning resonance has been fully resolved in
measurements extended down to 5 $cm^{-1}$ using a FIR laser. Also these
measurements reveal the existence of a zero frequency oscillator as has been
inferred from microwave studies. These observations are not germane to the
subject of this paper and will be published elsewhere.
\end{references}
\end{document}